\documentclass[prb, showpacs, twocolumn, aps, superscriptaddress, a4paper]{revtex4-1}
\usepackage{dcolumn, amssymb, amsmath, amsfonts, graphicx, latexsym, color, braket, subfigure}
\usepackage{epstopdf}
\usepackage{multirow}

\begin{document}

\title{Dissipative phase transitions in the fully-connected Ising model with $p$-spin interaction}
\author{Pei Wang}
\email{wangpei@zjnu.cn}
\affiliation{Department of Physics, Zhejiang Normal University, Jinhua 321004, People's Republic of China}
\author{Rosario Fazio}
\affiliation{Abdus Salam ICTP, Strada Costiera 11, I-34151 Trieste, Italy}
\affiliation{Dipartimento di Fisica, Universit\`a di Napoli ``Federico II'', Monte S. Angelo, I-80126 Napoli, Italy}
\thanks{On leave}

\date{\today}

\begin{abstract}
In this paper, we study the driven-dissipative p-spin models for $p\geq 2$. In thermodynamics limit,
the equation of motion is derived by using a semiclassical approach. The long-time asymptotic states
are obtained analytically, which exhibit multi-stability in some regions of the parameter space.
The steady state is unique as the number of spins is finite. But the thermodynamic limit
of the steady-state magnetization displays nonanalytic behavior somewhere inside the semiclassical multi-stable region.
We find both the first-order and continuous dissipative phase transitions.
As the number of spins increases, both the Liouvillian gap and magnetization
variance vanish according to a power law at the continuous transition.
At the first-order transition, the gap vanishes exponentially accompanied
by a jump of magnetization in thermodynamic limit. The properties of transitions depend
on the symmetry and semiclassical multistability, being qualitatively different among
$p=2$, odd $p$ ($p\geq 3$) and even $p$ ($p\geq 4$).
\end{abstract}

\maketitle

\section{Introduction}

Driven-dissipative systems are nowadays at the center of an intense
experimental and theoretical activity. Many different experimental platforms, 
from cavity arrays to BEC in cavities, just to mention some important examples,
have been realized leading to numerous interesting results. 
From the theoretical point of view, out of equilibrium phases and phase
transitions have properties with no (necessary) counterpart in equilibrium. 
Several reviews offer a wide perspective on the field~\cite{Houck2012,Sieberer2016,Hartmann2016,Noh2016}.

Many different models of driven-dissipative many-body dynamics have been scrutinized. Dissipation may radically alter the 
universality class of the transition itself, see e.g.~Refs.~[\onlinecite{Altman2015},\onlinecite{Magrebhi2016}].
Furthermore, the steady-state diagram may become much richer than its equilibrium counterpart. Many interesting situations have
been already investigated. The phase diagram of the Bose-Hubbard model in the presence of dissipative couplings was studied 
in~[\onlinecite{Diehl10,Wilson16,Rota19,Minganti18}]. The Ising model on a square lattice was shown to display a first-order 
or second-order phase transitions, depending on the form of the Lindblad operators~\cite{Weimer15,
Kshetrimayum17,Jin18}. The paramagnetic-ferromagnetic transition in the anisotropic XYZ-Heisenberg model 
was systematically studied by means of different approximation schemes~\cite{Lee13,Jin16,Biella18,Casteels18}.
Central-spin models were investigated as well, for example in~[\onlinecite{Kessler12}]. In the case of single-mode bosons,
the nature of transition was investigated, for example in~[\onlinecite{Casteels16}]. 
Dissipation causes the continuous time translational symmetry to be 
spontaneously broken, resulting in a time-crystal~\cite{Iemini17,Buca18,Lledo19,Seibold19}.
The dissipation here stabilizes a Floquet-time-crystal phase
in the central-spin model, as also confirmed by experiments on P-donor impurities in silicon~\cite{Sullivan18}. 
Similarly, the collective dissipation stabilizes the Floquet-time-crystal phase in the periodically-driven XY 
chain~\cite{Campeny19}. Finally, the quest for determining accurate calculation of the phase structure of open many-body 
systems has also prompted the development of accurate numerical tools~\cite{Weimer2019, Werner2016}

Topic of the present work is the study of the dissipative dynamics of a
fully connected model. The model describes a collection of 
spin-1/2 with  all-to-all couplings collectively coupled to an external bath.
In the presence of collective coupling to the environment~\cite{Iemini17,Hannukainen17} 
and transverse magnetic field (but no interaction in the Hamiltonian),
the total magnetization displays an everlasting oscillation in the thermodynamic limit, leading to 
a dissipative time-crystalline phase~\cite{Iemini17}. Phases with broken
time-translational invariance in dissipative systems appear also in several different 
models~\cite{Buca18,Tucker18,Gong17,Zhu19,Shammah18,Lledo19,Seibold19},
all of them essentially having long-range/collectively couplings
that admit mean-field like solutions.

The extension we will consider in this work, is to look at the properties of a
dissipative fully-connected spin model including the interaction among 
$p\geq 2$ spins. The Hamiltonian version of this model has a long history.
The $p$-spin model was initially introduced in a spin-glass 
context~\cite{Derrida80,Derrida81}, and has been under intensive investigations in the context of quantum annealing~\cite{Jorg09,Bapst12,Seoane12,
Matsuura17,Matsuura19,Ohkuwa18,Susa18,Yamashiro19,Passarelli18,
Passarelli19,Passarelli20,Passarelli20b,Re15,Wauters17}.
The dissipative dynamics of the $p$-spin model was explored in a recent
paper~\cite{Nava19} with Lindblad operators describing transitions between 
eigenstates of the Hamiltonian (hence chosen to guarantee thermalization).
The system under these conditions was shown to be 
trapped in a long-lived metastable state, realizing the so-called Mpemba
effect~\cite{Mpemba69}. In the present paper we consider collective 
Lindblad operators that lead to a non-equilibrium steady state. The resulting
phase diagram is very rich and strongly depends on the value of $p$.
Both first-order and continuous phase transitions are present, characterized
by different scaling behaviours of the Liouvillian gap and local observables.

The paper is organized as follows. We introduce the model in Sec.~\ref{sec:model}.
Sec.~\ref{sec:semimethod} focuses on the semiclassical approach and semiclassical phase diagram.
Sec.~\ref{sec:finiteN} discusses the steady-state properties at finite-$N$,
their scaling behavior and dissipative phase transitions. The real-time dynamics of magnetization
is discussed in Sec.~\ref{sec:finiteNdynamics}. Finally, Section~\ref{sec:sum} summarizes our conclusions.

\section{The model}
\label{sec:model}

The $p$-spin model consists of $N$ spins interacting through a site-independent interaction among $p$ of them. 
The corresponding  Hamiltonian reads
\begin{equation}\label{eq:pspinmodel}
\hat H = -\frac{J}{N^{p-1}} \sum_{j_1,j_2,\cdots,j_p=1}^N \hat \sigma_{j_1}^x\hat \sigma_{j_2}^x
\cdots \hat \sigma_{j_p}^x + h \sum_{j=1}^N \hat \sigma_j^z,
\end{equation}
where $\hat\sigma_j^\alpha$ with $\alpha=x,y,z$
are the Pauli matrices of the j-th spin. The couplings $J$ and $h$ are the interaction strength
and Zeeman field, respectively. We set $J= 1$ as the unit of energy throughout the paper.
The zero-temperature (ground state) phase diagram of the Hamiltonian~(\ref{eq:pspinmodel})  shows a transition at 
a critical value of the field $h$. For $p=2$, the  quantum phase transition is second-order, 
it becomes first-order~\cite{Jorg09} for $p>2$. The presence of the dissipative coupling radically changes the picture.

The dissipative generalization of model~\eqref{eq:pspinmodel} is introduced by coupling the system to an external 
reservoir. The dynamics of an open system is described by a suitable Master equation, obtained after integrating 
out the reservoirs' degrees of freedom. Here we consider a Markovian
dynamics with the evolution of the system's density matrix
governed by the Lindblad equation,  
\begin{equation}\label{eq:lindblad}
  \frac{ d\hat \rho}{dt} = -i \left[ \hat H, \hat \rho\right] + \frac{\kappa}{N} 
 \left[ 2\hat \sigma^- \hat \rho \hat \sigma^+ - \left\{\hat \rho , \hat \sigma^+ \hat \sigma^- \right\} \right].
\end{equation}
The Lindbladian, as well as the Hamiltonian, depends on the collective operators
$\hat \sigma^\pm= \sum_j \hat \sigma^\pm_j$ and $\hat \sigma^\pm_j
= \left(\hat \sigma^x_j \pm i \hat \sigma^y_j\right)/2$. The strength of the coupling to the environment is quantified by 
$\kappa>0$. In Eq.~\eqref{eq:lindblad}, the jump operator $\hat \sigma^-$  forces the spins to be aligned
in the negative $z$-direction (spin-down state). The Zeeman field introduces an energy difference between
the spin-up and spin-down states, while the $p$-spin interaction aligns the spins in the $x$-direction. 
If $p$ is even, Eq.~\eqref{eq:lindblad} has a $Z_2$ symmetry. We define the unitary transformation
$\hat U=\prod \otimes \hat \sigma^z_j$ which satisfies $\hat U^2=1$.
Indeed, $\hat U$ is a reflection transformation, which simultaneously
changes the sign of each spin's $x$- and $y$-components while keeping its
$z$-component invariant.  

The interplay between the  dissipative term and the Hamiltonian dynamics leads to a non-trivial steady state.
For $p=1$ the phase diagram was studied by Iemini et {al.}~\cite{Iemini17} and Hannukainen et {al.}~\cite{Hannukainen17}.
Here we extend the analysis to a generic $p$.

After an initial preparation  (for example in a pure state with all the spins
aligned in the same direction) the system is let evolve following 
Eq.~\eqref{eq:lindblad}. We are interested in the steady-state (long-time)
regime where the expectation values of observables are computed.
In thermodynamic limit, tuning the parameters $\kappa$ and $h$ may
lead to Dissipative Phase Transitions (DPTs) in the steady state. 
Two limits must be noticed here. One is the thermodynamic limit
$N\to \infty$, the other is $t\to \infty$ (dubbed the steady-state limit). The 
thermodynamic limit and the steady-state limit may not commute, i.e. $\displaystyle\lim_{t\to\infty}
\displaystyle\lim_{N\to\infty} \neq \displaystyle\lim_{N\to\infty} \displaystyle\lim_{t\to\infty}$.
For the problem we are interested here, by taking $N\to \infty$ first,
the dynamics can be studied by the semiclassical approach, to be described in the 
next Section. The corresponding semiclassical steady state will
be denoted as SSS, whose properties are discussed in Sec.~\ref{sec:semimethod}.
On the other hand, taking $t\to\infty$ first leads to the eigenstate of Liouvillian superoperator with zero eigenvalue.
The state obtained by taking $N\to\infty$ after $t\to\infty$ will be called the Liouvillian steady state (LSS),
whose properties are discussed in Sec.~\ref{sec:finiteN}.

\section{\label{sec:semimethod}Semiclassical approach}

We first consider the dynamics in the limit $N\to\infty$.
We define the operators $\hat s_\alpha=\sum_i \hat \sigma_i^\alpha/N$,
which satisfy the commutation relation $\left[ \hat s_\alpha,\hat s_\beta\right]
=2i \sum_\gamma \epsilon_{\alpha\beta\gamma} \hat s_\gamma/N$ with
$\epsilon_{\alpha\beta\gamma}$ denoting the antisymmetric tensor.
The magnetization components are the expectation values, i.e. $m_\alpha = 
\braket{\hat s_\alpha}=\textbf{Tr}\left[ \hat \rho \hat s_\alpha\right]$. To calculate the derivative
of $m_\alpha$ with respect to $t$, we substitute Eq.~\eqref{eq:lindblad} in
and notice $\braket{\hat s_\alpha\hat s_\beta}= \braket{\hat s_\alpha}
\braket{\hat s_\beta} $ as $N\to \infty$ in which limit $\hat s_\alpha$ and $\hat s_\beta$
are commutative~\cite{Iemini17}. The correlation vanishes in thermodynamic limit,
therefore, the equations of motion for $m_\alpha$ become
\begin{equation}\label{eq:meom}
\begin{split}
& \displaystyle\dot{m}_x = -2h m_y +\kappa m_x m_z, \\
& \displaystyle\dot{m}_y=2p \ m_x^{p-1} m_z +2hm_x+\kappa m_ym_z, \\
& \displaystyle\dot{m}_z=-2p \ m_x^{p-1} m_y -\kappa\left(m_x^2+m_y^2\right).
\end{split}
\end{equation}
$\left(m_x,m_y,m_z\right)$ satisfy a group of self-consistent
equations. At $t=0$, all the spins are supposed to be aligned
in a direction with azimuthal angles $\left(\theta,\phi\right)$. We
find the initial magnetization to be $m_x(0)=m \sin \theta \cos \phi$, $m_y(0)
=m\sin \theta \sin\phi$ and $m_z(0)=m \cos\theta$, where
$m=\sqrt{m_x^2+m_y^2+m_z^2}$ is the magnitude of magnetization.

It is easy to verify that $m$ is a constant of motion.
And Eq.~\eqref{eq:meom} keeps invariant
if we do the replacements $m_\alpha/m \to m_\alpha$, $m^{p-1} t \to t$,
$h/m^{p-1} \to h$ and $\kappa/m^{p-2} \to \kappa$. Therefore, we only
need to solve Eq.~\eqref{eq:meom} at $m=1$, and the other cases ($m\neq 1$)
can be mapped into it by rescaling the parameters.

As $m=1$, the vector $\left(m_x,m_y,m_z\right)$ is moving on a unit sphere
centered at the origin. We perform the stereographic map and map
the unit sphere into the $x$-$y$ plane. The map is defined as
\begin{equation}
\begin{split}
x= \frac{2m_x}{1-m_z}, \\
y=\frac{2m_y}{1-m_z},
\end{split}
\end{equation}
with the inverse map being
\begin{equation}
\begin{split}
m_x = \frac{4x}{4+x^2+y^2},\\
m_y=\frac{4y}{4+x^2+y^2}, \\
m_z= 1-\frac{8}{4+x^2+y^2}.
\end{split}
\end{equation}
After the stereographic map, the equations of motion become
\begin{equation}\label{eq:xyeom}
\begin{split}
\frac{dx}{dt} & = f(x,y) \\ & = -\kappa x-2hy-\frac{4^{p-1} p \ x^p y}{\left(x^2+y^2+4\right)^{p-1}}, \\
\frac{dy}{dt} & = g(x,y) \\ & = 2hx -\kappa y+\frac{1}{2} 
\frac{4^{p-1} p \ x^{p-1} \left(x^2-y^2-4\right)}{\left(x^2+y^2+4\right)^{p-1}}.
\end{split}
\end{equation}

The dynamics of the system in the $x$-$y$ plane can be well
understood by analyzing the properties of its fixed points which
are defined by $f(x,y)=g(x,y)=0$~\cite{Teschl12}. In our approach,
we first solve $\dot{m_x}=\dot{m_y}=\dot{m_z}=0$ to find
the fixed points on the unit sphere and then map them into the plane.
Obviously, the north pole (denoted by $\mathcal{N}$) and the south pole
(denoted by $\mathcal{S}$) are fixed points. The coordinates of 
$\mathcal{S}$ are $(m_x^{\mathcal{S}},m_y^{\mathcal{S}},m_z^{\mathcal{S}})
=(0,0,-1)$, while those of $\mathcal{N}$ are $(0,0,1)$.
Besides $\mathcal{N}$ and $\mathcal{S}$, there exist other fixed points
which are related to the solution of algebraic equation
\begin{equation}\label{eq:zeta}
\mu(\zeta) = \left(1+ \tilde\kappa \zeta \right)^p - \frac{p^2}{h^2} \zeta \left(1-\zeta\right)^{p-2}
=0
\end{equation}
in the domain $\zeta \in [0,1]$. Here we define $\tilde\kappa =\kappa^2/(4h^2)$.
When Eq.~\eqref{eq:zeta} has no root, only $\mathcal{N}$ and $\mathcal{S}$
are fixed points. Otherwise, Eq.~\eqref{eq:zeta} has
two roots denoted by $\zeta_1$ and $\zeta_2$, which satisfy
$0\leq \zeta_1 \leq \zeta_2 \leq 1$. Correspondingly, there exist four additional fixed points,
denoted by $\mathcal{P}_\pm$ and $\mathcal{Q}_\pm$.
As $p$ is odd, the coordinates of $\mathcal{P}_\pm$ are
\begin{equation}\label{eq:coorPQodd}
\begin{split}
& m_x^{\mathcal{P}_\pm} = \text{sign}(\mp h) \sqrt{(1-\zeta_{1})/(1+\tilde\kappa\zeta_{1})}, \\
& m_z^{\mathcal{P}_\pm}=\pm \sqrt{\zeta_1}.
\end{split}
\end{equation}
The coordinates of $\mathcal{Q}_\pm $ are obtained
by replacing $\zeta_1$ by $\zeta_2$ in Eq.~\eqref{eq:coorPQodd}.
As $p$ is even, the coordinates of $\mathcal{P}_\pm$ become
\begin{equation}\label{eq:coorPQeven}
\begin{split}
& m_x^{\mathcal{P}_\pm} =\pm \sqrt{(1-\zeta_1)/(1+\tilde\kappa \zeta_1)}, \\
& m_z^{\mathcal{P}_\pm}=-\text{sign}(h)\sqrt{\zeta_1}.
\end{split}
\end{equation}
And the coordinates of $\mathcal{Q}_\pm$ are obtained by replacing $\zeta_1$ by $\zeta_2$.
The $m_y$-coordinate is $m_y=\frac{\kappa}{2h} m_x m_z$ whether $p$ is odd or even.
Table~\ref{tab} summarizes the coordinates of $\mathcal{S}$, $\mathcal{P}_+$ and $\mathcal{P}_-$
for different $p$.

\begin{table}[b]
\renewcommand\arraystretch{2.0}
\begin{tabular}{| c | c | c | c |}
\hline
\multicolumn{2}{|c|}{} & Even $p$ & Odd $p$  \\
\hline
 \multirow{3}{*}{$\mathcal{P}_+$} & $m_x$ &
 $\sqrt{\displaystyle\frac{1-\zeta_{1}}{1+\tilde\kappa\zeta_{1}}}$ &
 $-\text{sign}(h) \sqrt{\displaystyle\frac{1-\zeta_{1}}{1+\tilde\kappa\zeta_{1}}}$ \\
\cline{2-4} & $m_y$ &
$- \sqrt{\displaystyle\frac{\tilde{\kappa}\zeta_1\left(1-\zeta_1\right)}{1+\tilde{\kappa}\zeta_1}}$ 
& $- \sqrt{\displaystyle\frac{\tilde{\kappa}\zeta_1
\left(1-\zeta_1\right)}{1+\tilde{\kappa}\zeta_1}}$  \\ \cline{2-4}
& $m_z$ & $-\text{sign}(h)\sqrt{\zeta_1}$ & $\sqrt{\zeta_1}$\\
\hline
 \multirow{3}{*}{$\mathcal{P}_-$} & $m_x$ &
 $-\sqrt{\displaystyle\frac{1-\zeta_{1}}{1+\tilde\kappa\zeta_{1}}}$&
 $\text{sign}(h) \displaystyle\sqrt{\frac{1-\zeta_{1}}{1+\tilde\kappa\zeta_{1}}}$ \\
\cline{2-4} & $m_y$ &
$\sqrt{\displaystyle\frac{\tilde{\kappa}\zeta_1
\left(1-\zeta_1\right)}{1+\tilde{\kappa}\zeta_1}}$ &
$- \sqrt{\displaystyle\frac{\tilde{\kappa}\zeta_1
\left(1-\zeta_1\right)}{1+\tilde{\kappa}\zeta_1}}$  \\ \cline{2-4}
& $m_z$ & $-\text{sign}(h)\sqrt{\zeta_1}$ & $- \sqrt{\zeta_1}$\\
\hline
 \multirow{3}{*}{$\mathcal{S}$} & $m_x$ & $0$ & $0$ \\
\cline{2-4} & $m_y$ & $0$ & $0$  \\ \cline{2-4}
& $m_z$ & $-1$ & $-1$\\
\hline
\end{tabular}
\caption{The magnetization components of $\mathcal{P}_+$,
$\mathcal{P}_-$ and $\mathcal{S}$ for different $p$.}\label{tab}
\end{table}

After the stereographic map, $\mathcal{S}$ is mapped into
the origin with the coordinates $\left(x_\mathcal{S},y_\mathcal{S}\right)=\left(0,0\right)$,
$\mathcal{N}$ is mapped into infinity, and
$\mathcal{P}_\pm$ and $\mathcal{Q}_\pm$ are mapped into somewhere in the
$x$-$y$ plane. Following the qualitative theory of differential equations,
we linearize Eq.~\eqref{eq:xyeom} around each fixed point.
In the vicinity of $\mathcal{P}_+$, Eq.~\eqref{eq:xyeom}
can be reexpressed as
\begin{equation}\label{eq:deltaeom}
\begin{split}
\frac{d}{dt} \left(\begin{array}{c} 
\Delta x \\ \Delta y \end{array} \right) =A_{\mathcal{P}_+}\left( \begin{array}{c} 
\Delta x \\ \Delta y \end{array} \right)
\end{split}
\end{equation}
with
\begin{equation}
\begin{split}\label{eq:deltaeom1}
A_{\mathcal{P}_+}= \left(
\begin{array}{cc} \left. \displaystyle\frac{\partial f}{\partial x}\right|_{\mathcal{P}_+} 
& \left. \displaystyle\frac{\partial f}{\partial y}\right|_{\mathcal{P}_+} \\
\left. \displaystyle\frac{\partial g}{\partial x}\right|_{\mathcal{P}_+}  & 
\left. \displaystyle\frac{\partial g}{\partial y}\right|_{\mathcal{P}_+} 
\end{array} \right) ,
\end{split}
\end{equation}
where $\Delta x = x-x_{\mathcal{P}_+}$ and $\Delta y = y- y_{\mathcal{P}_+}$ are
the coordinates relative to ${\mathcal{P}_+}$. In Eq.~\eqref{eq:deltaeom1},
the coefficient matrix has two eigenvalues. We find them to be
\begin{equation}\label{eq:lambdapm}
\lambda_\pm = \kappa m_z \pm \sqrt{(p-1)\kappa^2 - \frac{4h^2}{m_z^2} + 4h^2 (p-1)}.
\end{equation}
Here we have expressed $\lambda_\pm$ in terms of $(m_x,m_y,m_z)$,
whose values are given in Eqs.~\eqref{eq:coorPQodd} and~\eqref{eq:coorPQeven}.
Surprisingly, at the fixed points $\mathcal{P}_-$ or $\mathcal{Q}_\pm$,
the eigenvalues of coefficient matrix have the same expression as Eq.~\eqref{eq:lambdapm}.

Furthermore, in the vicinity of $\mathcal{S}$,
Eq.~\eqref{eq:xyeom} is linearized with the coefficient matrix being
\begin{equation}
\begin{split}
{A}_{\mathcal{S}}= \left(
\begin{array}{cc} -\kappa & -2h \\
2h - 4 I_p & -\kappa \end{array} \right),
\end{split}
\end{equation}
where $I_p$ equals $1$ for $p=2$ but $0$ for $p>2$.
The north pole $\mathcal{N}$ is mapped into infinity.
To analyze its properties, we directly linearize Eq.~\eqref{eq:meom} around
$(m^\mathcal{N}_x,m^\mathcal{N}_y,m^\mathcal{N}_z)=(0,0,1)$.
Replacing $m_z$ by $\sqrt{1-m_x^2-m_y^2}$, we find the linearized equation
to be $\left(\dot{m}_x,\dot{m}_y\right)^T= 
{A}_{\mathcal{N}} \left({m}_x,{m}_y\right)^T$, where
\begin{equation}\label{eq:eigenvaluesN}
\begin{split}
{A}_{\mathcal{N}}= \left(
\begin{array}{cc} \kappa & -2h \\
2h+4 I_p & \kappa \end{array} \right).
\end{split}
\end{equation}
The eigenvalues of $A_{\mathcal{S}}$ and $A_{\mathcal{N}}$ can be obtained easily.

The properties of a fixed point is exclusively determined by its eigenvalues
(see ref.~[\onlinecite{Teschl12}] for the classification of fixed points).
For $p=1$ the eigenvalues being purely imaginary indicates the
existence of limit cycle and then a time-crystalline phase~\cite{Iemini17}.
For $p\geq 2$ there exists no time-crystalline phase and the system
always relaxes towards a (time-independent) steady state.
In current model, if the real parts of $\lambda_+$ and $\lambda_-$ are
both negative, the fixed point is attractive, that is a point nearby
always moves towards the fixed point as $t$ increases and finally falls into
it as $t\to \infty$.  Except for a set of measure-zero  on the unit sphere,
starting from arbitrary $(m_x,m_y,m_z)$, one always ends at one of the
attractive fixed points in the limit $t\to \infty$.

\begin{figure*}[tbp]
\vspace{0.5cm}
\includegraphics[width=0.75\linewidth]{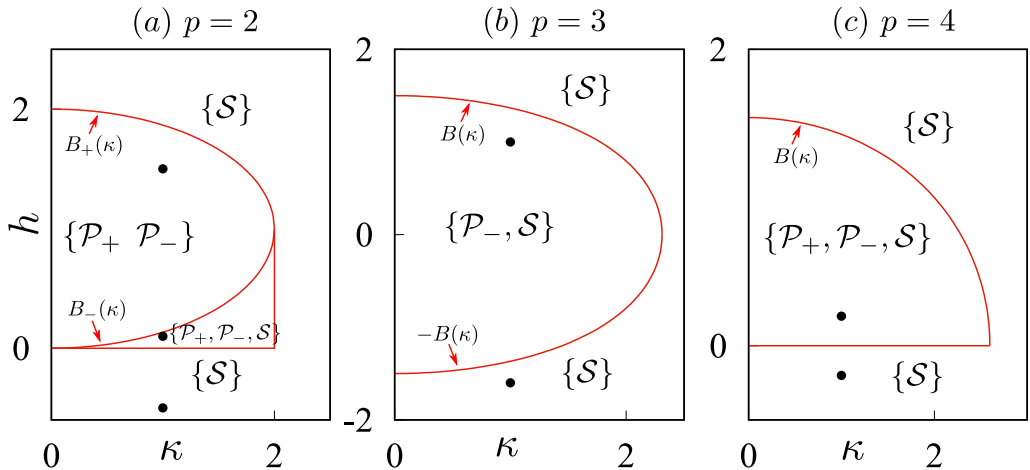}
\caption{(Color online) The semiclassical phase diagram in the
$\kappa$-$h$ half-plane for (a) $p=2$, (b) $p=3$ and (c) $p=4$.
The red solid lines are the borders between different semiclassical phases,
with $\mathcal{S}$ and $\mathcal{P}_\pm$ denoting the semiclassical steady states.
At the black dots, the real-time dynamics will be studied in Sec.~\ref{sec:finiteNdynamics}.}
\label{fig:phase}
\end{figure*}

To apply the semiclassical approach, we already take $N\to\infty$.
A physical state evolves into one of the attractive fixed points
if we take $t\to\infty$ after $N\to\infty$.
Therefore, the attractive fixed points are indeed the SSSs.
Next we check which of $\mathcal{N}$, $\mathcal{S}$, $P_\pm$ and $Q_\pm$
are attractive. We discuss the cases of
$p=2$, $p\geq 3$ being odd, and $p\geq 4$ being even, separately.

In the case $p=2$, two eigenvalues of $A_{\mathcal{N}}$
cannot both have negative real part, therefore, $\mathcal{N}$ is never attractive.
While $\mathcal{S}$ is attractive except for $0\leq \kappa \leq 2$
and $B_- \leq h \leq B_+$, where
\begin{equation}
B_\pm= 1\pm \sqrt{1-\kappa^2/4}.
\end{equation}
$\mathcal{P}_\pm$ are attractive fixed points if and only if
$0\leq \kappa \leq 2$ and $0\leq h \leq B_+$. But $\mathcal{Q}_\pm$ are never attractive.
It is easy to find the coordinates of $\mathcal{P}_\pm$ to be
$m_z^{\mathcal{P}_\pm}=-h/B_+$ and
$m_x^{\mathcal{P}_\pm}=\pm \sqrt{\frac{B_+}{2}-\frac{2h^2}{\kappa^2} B_-}$.
The $m_z$-coordinates of $\mathcal{P}_+$ and $\mathcal{P}_-$ are the same, but
their $m_x$-coordinates or $m_y$-coordinates have different signs.

The attractive fixed points (SSSs) in different regions of the $\kappa$-$h$ half-plane
are displayed in fig.~\ref{fig:phase}(a), in which $B_\pm(\kappa)$
are plotted in red lines. Three semiclassical phases are distinguished,
denoted by $\left\{\mathcal{S}\right\}$, $\left\{\mathcal{P}_+, \mathcal{P}_-,\mathcal{S}\right\}$
and $\left\{\mathcal{P}_+, \mathcal{P}_-\right\}$, respectively.
For $h<0$ or $h>B_+$, $\mathcal{S}$
is the unique attractive point. Starting from arbitrary initial state, $(m_x,m_y,m_z)$
always falls into the south pole in the steady limit, and
all the spins are aligned in the negative $z$-direction.
For $0<h<B_-$, there exist three attractive points, which are
$\mathcal{P}_\pm$ and $\mathcal{S}$. It depends on the initial condition whether a residue magnetization survives
in the $x$-direction or all the spins are aligned in the $z$-direction.
For $B_-<h<B_+$, the attractive points are $\mathcal{P}_\pm$. The steady state always has a nonzero
magnetization in the $x$-direction. Both the bistable and tri-stable phases
are located inside the area with $0\leq \kappa\leq 2$ and $0\leq h\leq 2$.
Furthermore, the bistable phase has a symmetric shape with respect to $h=1$.

For $p\geq 3$ and odd, $\mathcal{S}$ is attractive
but $\mathcal{N}$ is not in the whole parameter space.
Let us consider the polynomial $\mu(\zeta)$ defined in
Eq.~\eqref{eq:zeta}, which is a smooth function with the derivative of all orders being continuous.
The function $\dot\mu(\zeta) = d\mu/d\zeta$ increases monotonically in the interval $[0,1/(p-1)]$
and is definitely positive in the interval $[1/(p-1),1]$.
If $\dot\mu<0$ at $\zeta=0$, $\mu$ first decreases
with $\zeta$, reaches its minimum somewhere
and then increases with $\zeta$. If $\dot \mu \geq 0$ at $\zeta=0$,
$\mu$ increases monotonically with $\zeta$ in the domain $[0,1]$.
Since $\mu$ at the end points $\zeta=0,1$ are positive,
$\mu(\zeta)=0$ has either no root or
two roots (two roots can be the same).
If $\mu(\zeta)$ has roots, it must have a minimum in the domain $[0,1]$.
Suppose that the minimum is at $\zeta=\zeta_c$
with $\dot\mu(\zeta_c)=0$, the two roots then
satisfy $\zeta_1\leq \zeta_c\leq \zeta_2$.
For a given $\tilde\kappa$, $\mu(\zeta)$ is definitely positive for sufficiently large $h^2$
(check $\mu(\zeta)$ in the limit $h^2\to \infty$) and decreases
with $h^2 $ at every $\zeta$. There exists a critical value $h_c^2$
so that $\mu(\zeta)$ has no root for $h^2>h_c^2$, has
two same roots for $h^2=h_c^2$, but
has two different roots for $h^2<h_c^2$. At $h^2=h_c^2$,
one must have $\mu(\zeta_c) =0$. The two equations $\dot\mu(\zeta_c)=0$
and $\mu(\zeta_c)=0$ together determine $\zeta_c$ and $h_c^2$, which read
\begin{equation}\label{eq:zetahc}
\begin{split}
& \zeta_c= \frac{(p-1)(\tilde \kappa +1) -\sqrt{(p-1)^2\left(\tilde \kappa+1\right)^2
-4\tilde \kappa}}{2\tilde \kappa}, \\ &
h_c^2= \frac{p^2 \zeta_c\left(1-\zeta_c\right)^{p-2}}{\left(1+\tilde \kappa \zeta_c\right)^p}.
\end{split}
\end{equation}

At $h^2=h_c^2$, the roots of $\mu(\zeta)$ are $\zeta_1=\zeta_2=\zeta_c$.
For $h^2<h_c^2$, the roots of $\mu(\zeta)$ satisfy $\zeta_1< \zeta_c< \zeta_2$.
Eqs.~\eqref{eq:coorPQodd} and~\eqref{eq:coorPQeven} tell us
$\left(m_z^{\mathcal{P}_\pm}\right)^2=\zeta_1$ and
$\left(m_z^{\mathcal{Q}_\pm}\right)^2=\zeta_2$.
And Eq.~\eqref{eq:lambdapm} gives the eigenvalues $\lambda_\pm$ in terms of $m_z$.
We then express $\displaystyle\frac{\lambda_\pm}{2\left|h\right|}$
at the fixed points $\mathcal{P}_\pm$ ($\mathcal{Q}_\pm$)
in terms of $\zeta_1$ ($\zeta_2$).
By using Eq.~\eqref{eq:zetahc} and the relation $\zeta_1< \zeta_c< \zeta_2$,
we find that $\text{Re}(\lambda_\pm)$ at
$\mathcal{Q}_\pm$ or $\mathcal{P}_+$ can never be both negative,
but $\text{Re}(\lambda_\pm)$ at $\mathcal{P}_-$ are both negative. Therefore, $\mathcal{P}_-$ is
an attractive fixed point if and only if $h^2<h^2_c$.

In Eq.~\eqref{eq:zetahc}, $h_c^2$ is expressed in terms of $\tilde \kappa=\kappa^2/(4h_c^2)$.
For given $\kappa$, we solve Eq.~\eqref{eq:zetahc} to obtain $h_c^2$. Unfortunately, an explicit expression
of $h_c^2$ in terms of $\kappa$ is inaccessible. But it is not difficult to
see that there exists a critical $\kappa_c$.
For $\kappa>\kappa_c$, no $h_c^2$ satisfies Eq.~\eqref{eq:zetahc}.
For $\kappa\leq \kappa_c$, there exists a unique $h_c>0$ satisfying
Eq.~\eqref{eq:zetahc}, which will be denoted by $B$ from now on, or
\begin{equation}
B(\kappa)= \sqrt{h_c^2\left(\tilde{\kappa}\right)}.
\end{equation}
$B$ decreases with $\kappa$, reaching zero at $\kappa=\kappa_c$.

For $-B(\kappa) \leq h \leq B(\kappa)$, both $\mathcal{P}_-$ and $\mathcal{S}$
are attractive points. But for $\left |h\right| > B(\kappa)$, only
$\mathcal{S}$ is the attractive point. Fig.~\ref{fig:phase}(b)
displays the semiclassical phase diagram at $p=3$.
The $\kappa$-$h$ half-plane is divided into two  phases, denoted by $\left\{\mathcal{S}\right\}$ and 
$\left\{\mathcal{P}_-,\mathcal{S}\right\}$.
The $h$-axis and the curves $h=\pm B(\kappa)$
surrounds the bistable phase, in which the steady state has either finite
or vanishing magnetization in the $x$-direction, depending on the initial state.
The bistable phase has a symmetric shape with respect to $h=0$.
Outside the bistable phase, all the spins are aligned in the $z$-direction.
It is worth mentioning that the $Z_2$ symmetry is explicitly broken at $p=3$.
As a consequence, only $\mathcal{P}_-$ is a SSS, but $\mathcal{P}_+$ is not.

For $p\geq 4 $ and even, $\mathcal{S}$ is attractive
but $\mathcal{N}$ is not in the whole parameter space.
The above statements about the roots of $\mu(\zeta)$ still stand, including
Eq.~\eqref{eq:zetahc}. But the coordinates of $\mathcal{P}_\pm$ are different,
so is the phase diagram. Fig.~\ref{fig:phase}(c) shows the semiclassical phase diagram
of $p=4$. Two phases are distinguished, denoted by
$\left\{\mathcal{S}\right\}$ and $\left\{\mathcal{P}_+, \mathcal{P}_-,\mathcal{S}\right\}$.
For $h<0$ or $h>B(\kappa)$, the unique attractive point is $\mathcal{S}$.
For $0 \leq h \leq B(\kappa)$, there exist three attractive points,
which are $\mathcal{P}_\pm$ and $\mathcal{S}$.
The red lines surround the tri-stable phase, in which
the magnetization of SSS in the $x$-direction is zero or not,
depending on $\left(\theta,\phi\right)$. The tri-stable phase only exists for $h>0$.

\section{\label{sec:finiteN} Finite $N$}

\subsection{Method}

The semiclassical approach described before cannot be used to understand the finite-$N$ scaling behaviour. This regime is of 
particular importance if one would like to understand the critical properties of the Lindblad operator close to the DPTs. Moreover, 
on general grounds, we do expect that the multi-stability discussed in the previous Section will disappear at finite $N$ due to the 
tunnelling between the different semiclassical states.

For the present problem we can use the permutation symmetry to reduce the dimension
of Hilbert space from $2^N$ to $N+1$. Both the Lindblad equation~\eqref{eq:lindblad}
and initial state keep invariant under the exchange of arbitrary two spins ($\hat \sigma^\alpha_i \leftrightarrow
\hat \sigma^\alpha_j$). We then define the equal-weight basis (Dicke basis) as~\cite{Sciolla11}
\begin{equation}
\ket{s} = \frac{1}{\sqrt{C^{N(s+1/2)}_N}}\sum_{\sum_{j=1}^N \sigma_j^z = 2sN} 
\ket{\sigma_1^z, \sigma_2^z ,\cdots, \sigma_N^z},
\end{equation}
where $\sigma_j^z= \pm 1$ denotes the up and down states of the jth spin, respectively,
and $-N\leq 2sN\leq N$ is an integer. $s$ denotes
the averaged magnetization in the $z$-direction, which has $N+1$ different values,
that is $s=-\frac{1}{2}, -\frac{1}{2}+\frac{1}{N}, \cdots, \frac{1}{2}-\frac{1}{N},
\frac{1}{2}$. In the normalization factor, $C^{N(s+1/2)}_N$ is the binomial coefficient.
The equal-weight basis $\left\{\ket{s}\right\}$ generates a $(N+1)$-dimensional
subspace of the Hilbert space. Correspondingly, $\left\{\ket{s}\bra{s'}\right\}$ generates
a $(N+1)^2$-dimensional vector space of density matrix.
The evolution of $\hat\rho$ is limited in this space due to the permutation symmetry.

Therefore, $\hat \rho$ can be expressed as
$\hat \rho= \sum_{s,s'} \rho_{s,s'}\ket{s}\bra{s'}$. From now on, we
call $\rho_{s,s'}$ the density matrix. The Lindblad equation translates into
\begin{equation}
\begin{split}\label{eq:finiteNlind}
& \frac{1}{N}\dot{\rho}_{s,s'}(t) =  \left[ 2ih(s'-s)- \kappa \left(f^2_{s'}
+f^2_s\right) \right] \rho_{s,s'} \\ 
& + 2 \kappa f_{s+1/N} \ f_{s'+1/N} \ \rho_{s+1/N,s'+1/N} \\ &
+ iJ \left[ \left(\hat X_s \right)^p \rho_{s,s'} -\left(\hat X_{s'} \right)^p \rho_{s,s'} \right],
\end{split}
\end{equation}
where $f_s=\sqrt{\left(\frac{1}{2}+s\right)\left(\frac{1}{2}-s+\frac{1}{N}\right)}$
is a function of $s$. $\hat X_s$ and $\hat X_{s'}$ are linear operators
acting on the first and second arguments of $\rho_{s,s'}$, respectively.
$\hat X_s$ is defined by $\hat X_s \rho_{s,s'}=f_{s+1/N}\rho_{s+1/N,s'} +f_s \rho_{s-1/N,s'}$.
Similarly, $\hat X_{s'}$ keeps the first argument of $\rho_{s,s'}$ invariant
but changes its second argument.

Since the initial state is a pure state with all the spins aligned
in the direction $\left(\theta,\phi\right)$, its density matrix is
\begin{equation}
\begin{split}
\rho_{s,s'}(0)= & \left(\cos \frac{\theta}{2}\right)^{N(1+s+s')}\left(\sin\frac{\theta}{2}\right)^{N(1-s-s')}
e^{i\phi (s'-s)N} \\ & \sqrt{C^{N(s+1/2)}_N C_N^{N(s'+1/2)}}.
\end{split}
\end{equation}
Given $\rho_{s,s'}(0)$, Eq.~\eqref{eq:finiteNlind} can be solved numerically.
Because the dimension of density matrix is $N+1$, we can easily obtain
the solution for $N \leq 100$.  

After obtaining $\rho_{s,s'}(t)$, we can study the time-dependent magnetization components,
defined as $m_\alpha=\textbf{Tr}\left[\hat \rho \hat \sigma^\alpha\right]/N$.
Moreover, we define the variance of magnetization in the $x$-direction
to be $C_{x}=\displaystyle\frac{1}{N^2}
\textbf{Tr}\left[\hat \rho \left(\hat \sigma^x\right)^2 \right]$. 
These observables are connected to the density matrix by
\begin{equation}
\begin{split}\label{eq:mob}
m_x =&  \sum_s \rho_{s,s-1/N} f_s + \rho_{s,s+1/N} f_{s+1/N} \\
m_y =&  \sum_s i \rho_{s,s-1/N}f_s -i \rho_{s,s+1/N} f_{s+1/N} \\
m_z = & \sum_s 2 s \rho_{s,s} \\
C_x=&  \sum_s  \rho_{s,s-2/N} f_s f_{s-1/N} + \rho_{s,s} 
\left(f_s^2+f^2_{s+1/N} \right) \\ & + \rho_{s,s+2/N} f_{s+1/N}f_{s+2/N}.
\end{split}
\end{equation}

We re-express Eq.~\eqref{eq:finiteNlind} as
\begin{equation}\label{eq:finiteNmatrixlind}
\dot{\rho}_{s,s'} = \sum_{r,r'} \mathcal{L}_{s,s'; r,r'} \rho_{r,r'}, 
\end{equation}
where $\mathcal{L}$ is the Liouvillian matrix with $(s,s')$ denoting
its row index and $(r,r')$ denoting its column index. The dynamics of the system can be characterized by the eigenvalues and
eigenvectors of $\mathcal{L}$. The eigenvalues of $\mathcal{L}$
are denoted by $E_n$ with $n=1, 2,\cdots, (N+1)^2$.
$E_n$ either is a real number, or appears in conjugate pairs.
We rearrange the eigenvalues so that $\text{Re}(E_1)\geq \text{Re}(E_2) \geq \cdots$.
The eigenvalue with the largest real part must be $0$, that is $E_1=0$.
$E_2$ is the eigenvalue with the second largest real part.
$\Delta=-\text{Re}E_2> 0$ is  called the Liouvillian gap,
being always finite at finite $N$.
The eigenvectors are denoted by $\rho^{(1)},\rho^{(2)} \cdots$.

The initial density matrix can be decomposed as
\begin{equation}\label{eq:inicondlind}
\rho_{s,s'}(0)= \rho_{s,s'}^{(1)}+\sum_{n=2}^{(N+1)^2} K_n \rho^{(n)}_{s,s'}.
\end{equation}
Now the solution of Eq.~\eqref{eq:finiteNmatrixlind}
can be expressed as $\rho_{s,s'}(t)=  \rho_{s,s'}^{(1)}+
\sum_{n=2}^{(N+1)^2} e^{t E_n}K_n \rho^{(n)}_{s,s'}$. At finite $N$ we always have $\text{Re}E_n<0$
for $n\geq 2$, we then obtain
\begin{equation}
\displaystyle\lim_{t\to\infty} \rho_{s,s'}(t) = \rho^{(1)}_{s,s'}.
\end{equation}
The steady density matrix can be directly obtained
by diagonalizing $\mathcal{L}$ without solving a differential equation.
The expectation values of observables are computed
by using $\rho^{(1)}_{s,s'}$ and Eq.~\eqref{eq:mob}.
Because $\rho^{(1)}_{s,s'}$ depends on $N$, we denote the
observables obtained in this way as $m_\alpha^{(N)}$ and $C_x^{(N)}$.
Since the LSS is defined by taking $N\to\infty$ after $t\to\infty$,
the observables in LSS must be expressed as
\begin{equation}
\begin{split}
& m_\alpha^{\infty} = \displaystyle\lim_{N\to\infty} m_\alpha^{(N)}, \\
& C_x^{\infty}= \displaystyle\lim_{N\to\infty} C_x^{(N)}.
\end{split}
\end{equation}
It is worth emphasizing the different symbols that we use for LSS and SSS.
The magnetizations in the former are denoted by $m_\alpha^{\infty}$,
while those in the latter are denoted by $m_\alpha^{\mathcal{S}}$ or $m_\alpha^{\mathcal{P}_\pm}$.
Different from the SSSs, the LSS is unique for given $\left(\kappa,h\right)$ and $p$.

At finite $N$, the observables ($m_\alpha^{(N)}$ and $C_x^{(N)}$) are continuous functions of $h$ and $\kappa$.
But in the limit $N\to\infty$, the observables in LSS
display nonanalytic behavior somewhere inside the semiclassical multistable region.
Next we discuss these nonanalytic behaviors for $p=2$, $p=3$ and $p=4$ separately.

\subsection{\label{sec:p2}$p=2$}

\begin{figure}[tbp]
\vspace{0.5cm}
\includegraphics[width=0.9\linewidth]{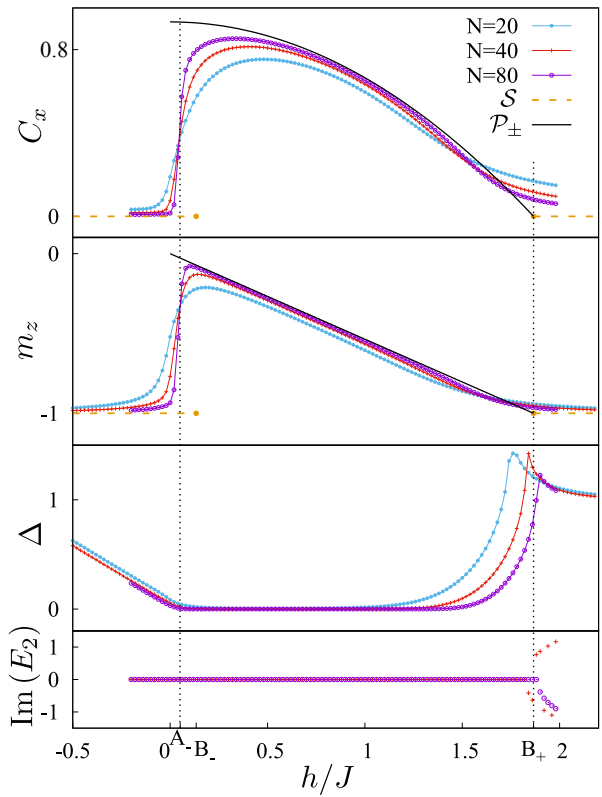}
\caption{(Color online) From top to bottom, we plot the variance in the $x$-direction,
magnetization in the $z$-direction, Liouvillian gap and $\text{Im}\left(E_2\right)$ as a function of $h$.
Different dot types are for different $N$. The black solid lines are
$m_z^{\mathcal{P}_\pm}$ or $C_x^{\mathcal{P}_\pm}$,
while the orange dashed lines are $m_z^\mathcal{S}$ or $C_x^{\mathcal{S}}$.
All the panels share the same legends.
$B_-= 0.134$ is the border between tri-stable and bistable semiclassical phases.
$B_+=1.866$ is the border between bistable and trivial phases, where
a continuous transition happens.
A first-order transition happens at $A_-=0.052$. The black dotted lines indicate the
location of $A_-$ and $B_+$.}\label{fig:hfuncp2}
\end{figure}

The Lindblad equation has a reflection symmetry as $p$ is even, at the same time,
the steady state is unique at finite $N$. Therefore, the steady state
keeps invariant under the reflection transformation $\hat U$.
But $\hat U$ acting on a state changes the signs of $m_x^{(N)}$ and $m_y^{(N)}$.
It is possible only if $m_x^{(N)}$ and $m_y^{(N)}$ are both zero.
Naturally, $m_x^\infty$ and $m_y^\infty$ in LSS must be zero.
The steady-state magnetization has only $z$-component, which is apparently different
from the semiclassical result. On the other hand, even if $C_x=m_x^2$ is guaranteed in semiclassical approach due to
the lack of correlation, it is not the case at finite $N$ or in the LSS.
$C_x^{(N)}$ is finite while $m_x^{(N)}$ is zero.
We find that $m_z^\infty$ and $C_x^\infty$ as a function of $h$ display nonanalyticity
for $0<\kappa<2$. Next we choose $\kappa=1$ to demonstrate their properties.

From top to bottom in Fig.~\ref{fig:hfuncp2}, we plot the variance in the $x$-direction,
magnetization in the $z$-direction, Liouvillian gap and $\text{Im}\left(E_2\right)$
as a function of $h$ for different $N$. The observables in SSSs,
i.e. $m^{\mathcal{S}}_z$, $m^{\mathcal{P}_\pm}_z$,
$C_x^{\mathcal{S}}=\left(m^{\mathcal{S}}_x\right)^2$
and $C_x^{\mathcal{P}_\pm}=\left(m^{\mathcal{P}_\pm}_x\right)^2$,
are plotted as a comparison (orange dashed and black solid lines).
We mark the positions of $B_-$ and $B_+$ on the $h$-axis. Recall that, the set of SSSs
is $\left\{\mathcal{S},\mathcal{P}_+, \mathcal{P}_-\right\}$ as $h\in [0,B_-]$,
$\left\{\mathcal{P}_+, \mathcal{P}_-\right\}$ as $h\in [B_-, B_+]$, but $\left\{\mathcal{S}\right\}$ otherwise.

At finite $N$, $m_z^{(N)}$ and $C_x^{(N)}$ are both smooth functions of $h$,
and $\Delta $ is finte. For $h< 0$, $\Delta$ drops linearly with increasing $h$.
The Liouvillian gap almost vanishes at $h=A_-\approx 0.052$. On the other hand, $m_z^{(N)}(h)$
of different $N$ cross each other at $h=A_-$. $m_z^{(N)}$
is close to $m_z^\mathcal{S}=-1$ as $h$ is much smaller than $A_-$, but it increases
abruptly at $h=A_-$ and reaches almost $m_z^{\mathcal{P}_\pm}$.
Similar behavior is seen in $C_x^{(N)}$, which is close to zero for
$h \ll A_-$ but increases abruptly at $h=A_-$ and
reaches almost $C_x^{\mathcal{P}_\pm}$. $C_x^{(N)}(h)$ of different $N$
also cross each other at $h=A_-$,
which helps us to determine the value of $A_-$. As $N$ increases from $20$ to $80$,
the change of $m_z^{(N)}$ or $C_x^{(N)}$ at $h=A_-$ becomes sharper.
We guess that $m_z^\infty$ and $C_x^\infty$ should be discontinuous at $h=A_-$,
which signals a first-order phase transition.

For $h>A_-$, the gap $\Delta$ stays small in the interval $(A_-,1.5)$, but increases
abruptly around $h=B_+$. While $m_z^{(N)}$ and $C_x^{(N)}$
drops continuously with increasing $h$. In the vicinity of $h=B_+$,
$m_z^{(N)}$ drops back close to $m_z^\mathcal{S}=-1$ and $C_x^{(N)}$ drops back close to $C_x^\mathcal{S}=0$.
Notice that the fixed points $\mathcal{P}_\pm$ meet $\mathcal{S}$ (the south pole) at $h=B_+$
(see e.g. the orange dashed and black solid lines).
For $h\in [A_-,B_+]$, as $N$ increases from $20$ to $80$, $m_z^{(N)}$ ($C_x^{(N)}$)
goes closer to $m_z^{\mathcal{P}_\pm}$ ($C_x^{\mathcal{P}_\pm}$).
Especially, at $h=B_+$, $m_z^{(N)}$ or $C_x^{(N)}$ decrease towards
$m_z^{\mathcal{P}_\pm}= m_z^{\mathcal{S}}=-1$ or $C_x^{\mathcal{P}_\pm}
=C_x^{\mathcal{S}}=0$ with increasing $N$, respectively.
The observables are smooth functions of $h$ at finite $N$ so that
$m_z^{(N)}$ ($C_x^{(N)}$) cannot really reach $-1$ ($0$).
But it is reasonable to guess $m_z^\infty=-1$ and $C_x^\infty=0$ 
for $h\geq B_+$. If this is true, $m_z^\infty(h)$
and $C_x^\infty(h)$ are nonanalytic at $h=B_+$ where a continuous
phase transition happens. 

In the vicinities of $h=A_-$ or $h=B_+$, we see that $\text{Im}(E_2)$ vanishes
(fig.~\ref{fig:hfuncp2} the bottom panel). For large enough $N$ (e.g. $N=80$),
$\text{Im}(E_2)$ vanishes in the whole multistable phase ($0<h<B_+$).

\begin{figure}[tbp]
\vspace{0.5cm}
\includegraphics[width=0.9\linewidth]{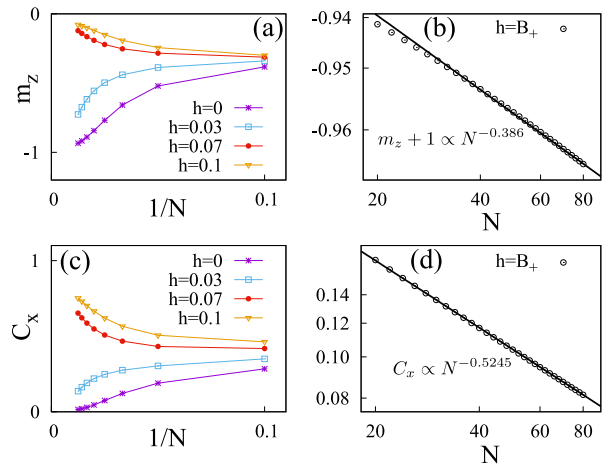}
\caption{(Color online) $m_z^{(N)}$ and $C_x^{(N)}$ as a function of $1/N$
are plotted in panels (a) and (c) for $h$ in the vicinity
of $A_-=0.052$, respectively. Panels (b) and (d) plot
$m_z^{(N)}$ and $C_x^{(N)}$ at $h=B_+$ with circles, respectively.
And the solid lines are fits to data. Both axes of~(b) or~(d)
are in logarithmic scale.}\label{fig:scalingobp2}
\end{figure}

To confirm the nonanalyticity of observables at $h=A_-$ and $h=B_+$,
we do a scaling analysis. Figs.~\ref{fig:scalingobp2}(a)
and~\ref{fig:scalingobp2}(c) plot $m_z^{(N)}$ and $C_x^{(N)}$ as a function of $1/N$
in the vicinity of $h=A_-$, respectively. As $1/N \to 0$ ($N\to\infty$),
the observables at $h<A_-$ ($h=0$ and $h=0.03$) decrease, but those
at $h>A_-$ ($h=0.07$ and $h=0.1$) increase. This bifurcation
is a clear signature of $m_z^\infty$ and $C_x^\infty$ being discontinuous at $h=A_-$.
Fig.~\ref{fig:scalingobp2}(b) and fig.~\ref{fig:scalingobp2}(d) plot the observables
as a function of $N$ in logarithmic scale at $h=B_+$.
Both $m_z^{(N)}+1$ and $C_x^{(N)}$ decay with increasing $N$,
and they decay according to a power law. Their scaling behavior at large $N$
can be approximately expressed as
\begin{equation}\label{eq:scalingobp2}
\begin{split}
&C^{(N)}_x(B_+)
\propto N^{-0.5245}, \\
& m^{(N)}_z(B_+)+1 \propto N^{-0.386}.
\end{split}
\end{equation}
In the limit $N\to\infty$, we have $C_x^\infty=0$ and $m_z^\infty=-1$.
The magnetization variance in the $x$-direction vanishes at $h=B_+$.
The scaling law~\eqref{eq:scalingobp2} is reminiscent of what we see
at a continuous phase transition in ground states, indicating some similarity
between continuous DPT and quantum phase transition.

\begin{figure}[tbp]
\vspace{0.5cm}
\includegraphics[width=0.9\linewidth]{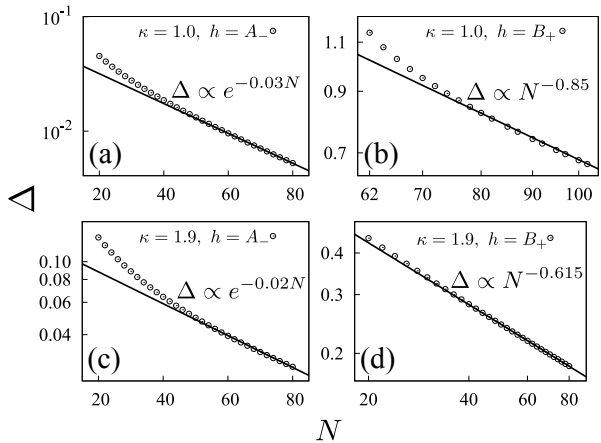}
\caption{(Color online) The Liouvillian gap as a function of $N$
is plotted with circles for $\kappa=1.0, 1.9$ and $h=A_-, B_+$. The
solid lines are fits to data. Notice that the $N$-axis of~(a) and~(c)
is displayed in normal scale but that of~(b) and~(d) is displayed in logarithmic
scale.}\label{fig:p2Escale}
\end{figure}

It was argued that the closed Liouvillian gap is a necessary
condition of the nonanalyticity in LSS~\cite{Kessler12}. This gap is finite at finite $N$,
and only closes in the limit $N\to\infty$. We then study how $\Delta$ changes
with increasing $N$ at $h=A_-$ and $h=B_+$. Fig.~\ref{fig:p2Escale} plots $\Delta$ as a function
of $N$. We not only show the gap at $\kappa=1.0$, but also show the gap at $\kappa=1.9$
for a comparison. The behavior of observables and gaps at $\kappa =1.9$ is similar to that at $\kappa=1.0$.
Ref.~[\onlinecite{Casteels16}] argued that the Liouvillian gap vanishes
exponentially at a first-order transition. Fig.~\ref{fig:p2Escale}(a) and fig.~\ref{fig:p2Escale}(c)
display $\Delta$ in logarithmic scale at $h=A_-$.
For large $N$, $\Delta$ is approximately an exponential function of $N$
with the exponents being $-0.03N$ and $-0.02N$ for $\kappa=1.0$ and
$\kappa=1.9$, respectively. The exponential fit at $\kappa=1.0$ looks
better than that at $\kappa=1.9$, possibly because the gap at $\kappa=1.0$
is smaller. On the other hand, the Liouvillian gap vanishes according to a power law
at $h=B_+$. This is observed for both $\kappa=1.0$ and $\kappa=1.9$
(see fig.~\ref{fig:p2Escale}(b) and fig.~\ref{fig:p2Escale}(d)). When
$N$ is large, we find $\Delta \propto N^{-\nu}$
with $\nu \approx {0.85}$ ($\nu \approx{0.615}$) for $\kappa=1.0$
($\kappa=1.9$). The power-law vanishing of Liouvillian gap is related to the power-law
vanishing of $C_x^N$. The continuous transition at $h=B_+$ has different properties from
the first-order transition at $h=A_-$.

In short, we find a first-order DPT at $h=A_-(\kappa)$, which is characterized by
the discontinuity of $C_x^\infty$ and $m_z^\infty$.
But at $h=B_+(\kappa)$, the DPT is continuous with $C_x^\infty$ vanishing continuously and $m_z^\infty$ dropping
continuously to $-1$. For $A_-<h<B_+$, the LSS is nontrivial,
indicated by $C_x^\infty>0$ and $m_z^\infty\neq -1$.
In the nontrivial region of LSS, the magnetization in the $x$-direction
is zero but its variance is finite, indicating that the nontrivial state is
a mixture of $\mathcal{P}_+$ and $\mathcal{P}_-$. Due to the $Z_2$ symmetry,
there are equal probabilities for
the spins to be aligned in the positive or negative $x$-direction.
Therefore, the averaged magnetization is zero but the averaged
square of magnetization is finite. We have $C_x^\infty\neq \left(m_x^\infty\right)^2$
in the nontrivial region. The correlation survives in thermodynamic limit.

We notice $0<A_-<B_-$. The first-order DPT happens inside the
tri-stable semiclassical phase in which $\mathcal{P}_\pm$ are different
from $\mathcal{S}$. But the continuous DPT happens at the border
between the bistable phase $\left\{\mathcal{P}_\pm\right\}$
and the phase $\left\{\mathcal{S}\right\}$.
At this border, the fixed points $\mathcal{P}_\pm$ move to $\mathcal{S}$,
explaining why $C_x^\infty$ or $m_z^\infty$ are continuous here.

We emphasize that our numerical method only works at small $N$. For $N$ much larger
than $100$, it is hard to obtain the steady-state observables and $\Delta$ by diagonalizing
the Liouvillian superoperator, since the dimension of density-matrix space grows as $N^2$.
But there exists finite-size effect at small $N$. Our conclusions about $m_\alpha^\infty$,
$C_x^\infty$ or $\Delta^\infty$ may be seriously influenced by the finite-size effect.

\subsection{\label{sec:p3}$p=3$}

\begin{figure}[tbp]
\vspace{0.5cm}
\includegraphics[width=0.9\linewidth]{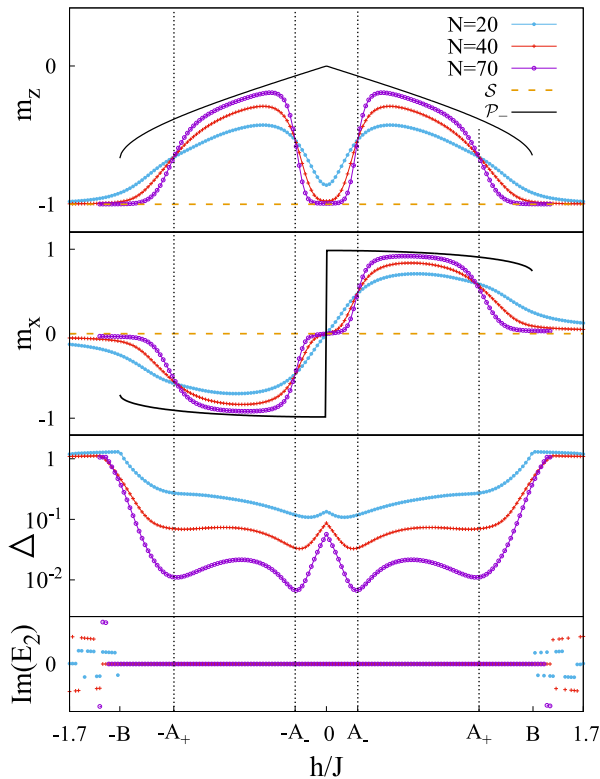}
\caption{(Color online) From top to bottom, we plot the magnetization
in the $z$-direction, magnetization in the $x$-direction, Liouvillian gap,
and imaginary part of $E_2$ as a function of $h$. Different dot colors and types are for different $N$.
The coordinates of $\mathcal{S}$ (orange dash) and
$\mathcal{P}_-$ (black solid) are plotted as comparison. All the panels share the same legends.
$\pm B= \pm 1.367$ are the boundary of bistable phase in which $\mathcal{P}_-$
exists. The first-order DPTs happen at $h=\pm A_+=\pm 1.01$ and $h=\pm A_-=\pm 0.207$.
The black dotted lines indicate their locations.}\label{fig:obp3}
\end{figure}

$Z_2$ symmetry is explicitly broken at $p=3$.
The properties of LSS are then significantly different from those at $p=2$.
Especially, the magnetizations in the $x$- or $y$-direction are not necessarily zero.
The magnetization as a function of $h$ displays nonanalyticity somewhere inside the bistable phase
for $\kappa$ being not very large. We choose $\kappa=1.0$ to demonstrate the behavior
of magnetization.

Fig.~\ref{fig:obp3} the top panels plot $m_z^{(N)}$ and $m_x^{(N)}$
as a function of $h$ for $N=20, 40$ and $70$.
The magnetizations of semiclassical states $\mathcal{S}$ and $\mathcal{P}_-$ are plotted in the same figure
for comparison. $\mathcal{P}_-$ exists only for $h\in[-B,B]$
where $\pm B$ are the borders between stable ($\{\mathcal{S}\}$) and bistable
($\{\mathcal{S}, \mathcal{P}_-\}$) phases.

$m_z^{(N)}(h)$ is an even function of $h$, but $m_x^{(N)}(h)$ is an odd function.
This fact can be explained by using the properties of $\hat \sigma^z$ and $\hat \sigma^x$
under the reflection transformation $\hat U$.
Within the interval $[-B,B]$, the curves $m_z^{(N)}(h)$ for different $N$
cross each other at $h=-A_+$, $- A_-$, $A_-$ and $A_+$. And the curves
$m_x^{(N)}(h)$ cross each other at exactly the same $h$. The positions
of $\pm A_+\approx \pm 1.01$ and $\pm A_-\approx \pm 0.207$ are
indicated by the black dotted lines in fig.~\ref{fig:obp3}.
At $h=\pm A_+, \pm A_-$, both $m_z^{(N)}(h)$ and $m_x^{(N)}(h)$
display abrupt changes. For $h<-A_+$, $m_z^{(N)}$ is close to
$-1$ and $m_x^{(N)}$ is close to $0$,
that is they are close to the coordinates of $\mathcal{S}$.
At $h=-A_+$, $m_z^{(N)}$ increases abruptly to $m_z^{\mathcal{P}_-}$,
while $m_x^{(N)}$ drops abruptly to $m_x^{\mathcal{P}_-}$. The second
transition happens at $h=-A_-$, where $m_z^{(N)}$
and $m_x^{(N)}$ go back to the coordinates
of $\mathcal{S}$. And similar transitions between $\mathcal{P}_-$
and $\mathcal{S}$ happen at $h=A_-$ and $A_+$.
As $N$ increases from $20$ to $70$, the changes of $m_x^{(N)}$ and $m_z^{(N)}$
become sharper. We then expect that the change becomes discontinuous
in thermodynamic limit.

\begin{figure}[tbp]
\vspace{0.5cm}
\includegraphics[width=0.9\linewidth]{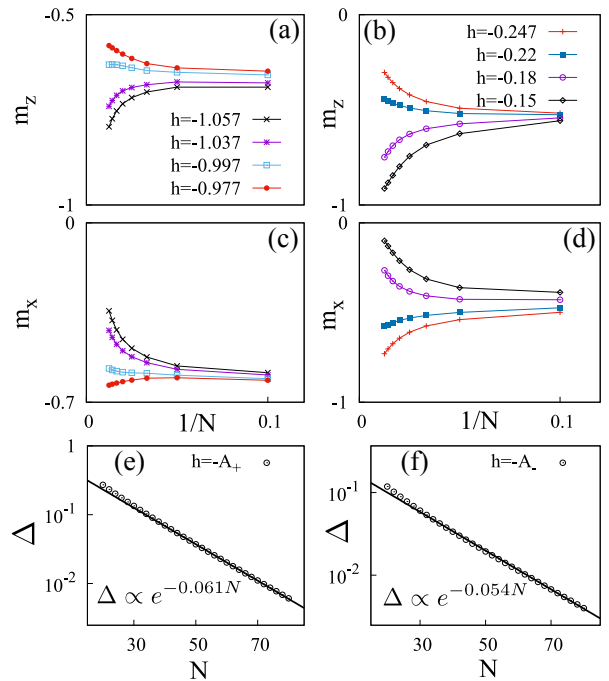}
\caption{(Color online) (a) $m_z^{(N)}$ vs $1/N$ in the vicinity
of $h=-A_+=-1.01$. (b) $m_z^{(N)}$ vs $1/N$ in the vicinity
of $h=-A_-=-0.207$. (c) $m_x^{(N)}$ vs $1/N$ in the vicinity
of $h=-A_+$. (d) $m_x^{(N)}$ vs $1/N$ in the vicinity
of $h=-A_-$. (a) and (c) share the same legends. (b)
and (d) share the same legends. (e) and (f) plot $\Delta$ vs $N$
with circles at $h=-A_+$ and $-A_-$, respectively,
with the solid lines being fits to data.}\label{fig:scalep3}
\end{figure}

Fig.~\ref{fig:obp3} the bottom panels plot $\Delta = -\text{Re}(E_2)$ and
$\text{Im}(E_2)$ as a function of $h$. Notice that $\Delta$
is displayed in logarithmic scale. It is clear that $\Delta(h)$ is an even function
and reaches its local minimum at $h=\pm A_+, \pm A_-$.
As $N$ increases, $\Delta$ at the transition points decays
towards zero, and is as small as $10^{-2}$ at $N=70$.
We also see that $\text{Im}(E_2)$ vanishes within the interval $[-B,B]$ for arbitrary $N$.
In other words, $\text{Im}(E_2)$ stays zero in the vicinity of $\pm A_+$ or $\pm A_-$,
similar to what happens in the case of $p=2$.

To confirm the vanishing gap and discontinuity of observables at $h=\pm A_+$ or $\pm A_-$,
we display their scaling behavior in fig.~\ref{fig:scalep3}. The panels~(a)
and~(c) are for $m_z^{(N)}$ and $m_x^{(N)}$ in the
vicinity of $h=-A_+$, respectively.
As $1/N$ goes to zero ($N\to\infty$), $m_z^{(N)}$
bends towards different directions for $h>-A_+$ ($h=-0.997, -0.977$)
and for $h<-A_+$ ($h=-1.057,-1.037$), so does $m_x^{(N)}$.
This indicates that $m_z^\infty$ or $m_x^\infty$ are discontinuous at $h=-A_+$.
The panels~(b) and~(d) show $m_z^{(N)}$ and $m_x^{(N)}$ in the
vicinity of $h=-A_-$, respectively. Similarly, the bifurcation indicates
that $m_z^\infty$ or $m_x^\infty$ are discontinuous at $h=-A_-$.
Because $m_z^\infty$ ($m_x^\infty$) is an even (odd) function, they must be also discontinuous
at $h=A_-$ and $A_+$.

Fig.~\ref{fig:scalep3}(e) and fig.~\ref{fig:scalep3}(f) display $\Delta$
as a function of $N$ at $h=-A_+$ and $h=-A_-$,
which perfectly fit the curves $e^{-0.061N}$ and $e^{-0.054N}$, respectively.
The Liouvillian gaps vanish exponentially at the critical points,
which is the feature of first-order transition.

As $p$ is odd, the $Z_2$ symmetry is broken.
$m_x^\infty$ is an odd function of $h$, while $m_z^\infty$ is an even function.
All the DPTs are first-order, characterized by the discontinuity
of $m_x^\infty$ and $m_z^\infty$. The first-order DPTs happen
at $h=\pm A_-$ and $\pm A_+$, which are located inside the bistable
phase in which $\mathcal{S}$ and $\mathcal{P}_-$ are two different SSSs.
LSS is one of the semiclassical states.
In the regions $\left| h\right| \leq A_-(\kappa)$ or $\left| h\right| \geq A_+(\kappa)$,
LSS is $\mathcal{S}$ with spins polarized in the negative $z$-direction.
For $A_- (\kappa)< \left| h\right| < A_+(\kappa)$,
LSS is $\mathcal{P}_-$ with finite magnetization in the $x$-direction.
Again, we would like to emphasize that our conclusions are based on the observation
at small $N$ limited by the numerical method.

\subsection{\label{sec:p4}$p = 4$}

\begin{figure}[tbp]
\vspace{0.5cm}
\includegraphics[width=0.9\linewidth]{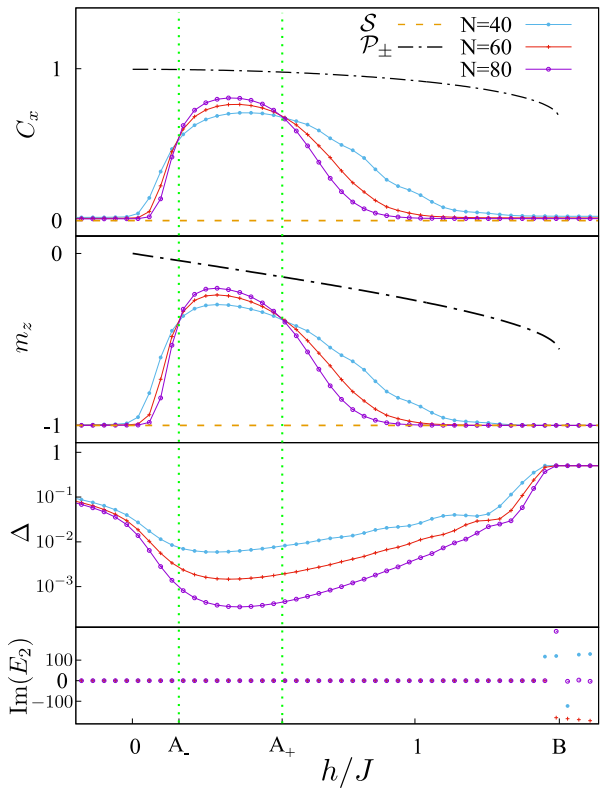}
\caption{From top to bottom, we plot the magnetization variance
in the $x$-direction, magnetization in the $z$-direction, Liouvillian gap,
and $\text{Im}(E_2)$ as a function of $h$. Different dot colors and types are for different $N$.
The magnetization and variance of $\mathcal{S}$ (orange dash) and
$\mathcal{P}_\pm$ (black dash-dotted) are plotted together
as comparison. All the panels share the same legends.
The system is in the tri-stable phase for $h\in [0,B]$ in which $\mathcal{P}_\pm$ exist.
The transition happens at $A_-\approx 0.165$ and $A_+\approx 0.53$.
The green dotted lines indicate the locations of $A_\pm$.}\label{fig:obp4}
\end{figure}

The $Z_2$ symmetry is present at $p=4$.
The symmetry forces $m_x^{(N)}=m_y^{(N)}=0$ and then $m_x^\infty=m_y^\infty=0$.
But the semiclassical phase diagram of $p=4$ is qualitatively different from that of $p=2$.
$m_z^\infty$ and $C_x^\infty$ then show different features. They display nonanalytic behavior
inside the tri-stable phase only for small enough $\kappa$ (e.g. $\kappa=0.5 J$).
But for large $\kappa$ (e.g. $\kappa=J$), as $N$ increases, $m_z^{(N)}$ always asymptotes $-1$ and
$C_x^{(N)}$ asymptotes $0$ for arbitrary $h$.
Next we focus on the nonanalytic behavior at $\kappa=0.5 J$.

We plot $C_x^{(N)}$ and $m_z^{(N)}$ as a function of $h$ for $N=40, 60$ and $80$
in fig.~\ref{fig:obp4} the top panels. The observables in the SSSs, i.e.
$m_z^{\mathcal{S}}$, $m_z^{\mathcal{P}_\pm}$, $C_x^{\mathcal{S}}$
and $C_x^{\mathcal{P}_\pm}$, are plotted together
for comparison. Remember that $\mathcal{P}_\pm$
exist only for $h\in[0,B]$ in which interval the system is tri-stable.

Within the interval $[0,B]$, the curves $C_x^{(N)}(h)$ ($m_z^{(N)}(h)$) for different $N$
cross each other at $h=A_-\approx 0.165$ and $A_+\approx 0.53$.
At $h=A_\pm$, $C_x^{(N)}(h)$ and $m_z^{(N)}(h)$
display abrupt changes. For $h<A_-$, $C_x^{(N)}$ ($m_z^{(N)}$) is close to $0$
($-1$), that is the coordinates of $\mathcal{S}$.
At $h=A_-$, $C_x^{(N)}$ ($m_z^{(N)}$) increases abruptly towards $C_x^{\mathcal{P}_\pm}$
($m_z^{\mathcal{P}_\pm}$). At the second critical point $h=A_+$,
$C_x^{(N)}$ ($m_z^{(N)}$) drops back to $0$ ($-1$).
As $N$ increases from $40$ to $80$, the changes of $C_x^{(N)}$ and $m_z^{(N)}$
become sharper. We expect that they are discontinuous in thermodynamic limit.

Fig.~\ref{fig:obp4} the bottom panels show $\Delta$ vs $h$ and
$\text{Im}(E_2)$ vs $h$. For $h\in[A_-,A_+]$, $\Delta$
decreases with increasing $N$, and is less than $10^{-3}$ for $N=80$.
We expect the gap to close at the transition points $h=A_\pm$ in thermodynamic limit.
$\text{Im}(E_2)$ vanishes in the vicinities of $A_\pm$ for arbitrary $N$,
similar to what happens in the cases of $p=2$ or $p=3$.

\begin{figure}[tbp]
\vspace{0.5cm}
\includegraphics[width=0.9\linewidth]{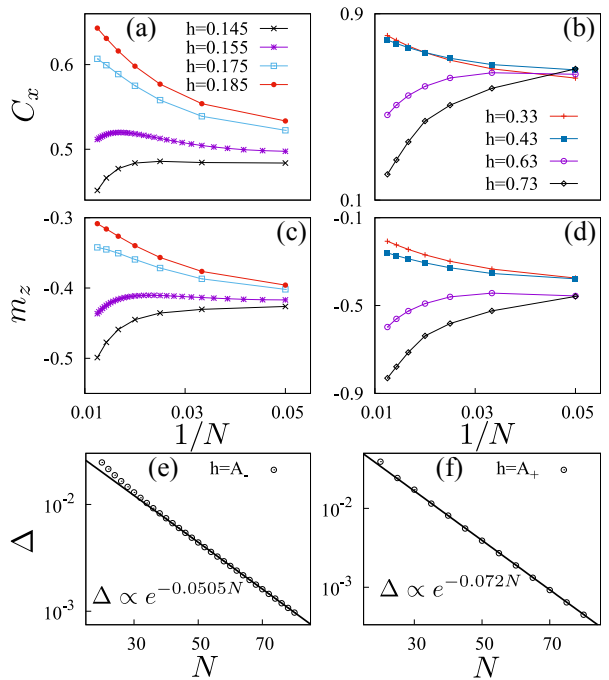}
\caption{(Color online) (a) $C_x^{(N)}$ vs $1/N$ in the vicinity
of $h=A_-=0.165$. (b) $C_x^{(N)}$ vs $1/N$ in the vicinity
of $h=A_+=0.53$. (c) $m_z^{(N)}$ vs $1/N$ in the vicinity
of $h=A_-$. (d) $m_z^{(N)}$ vs $1/N$ in the vicinity
of $h=A_+$. (a) and (c) share the same legends. (b)
and (d) share the same legends. (e) and (f) plot $\Delta$ vs $N$
with circles at $h=A_-$ and $A_+$, respectively,
and the solid lines are fits to data.}\label{fig:p4scaling}
\end{figure}
We confirm the vanishing gap and discontinuity of observables at $h=A_\pm$ by
doing a scaling analysis (see fig.~\ref{fig:p4scaling}). Fig.~\ref{fig:p4scaling}(a)
and fig.~\ref{fig:p4scaling}(c) plot $C_x^{(N)}$ and $m_z^{(N)}$ in the
vicinity of $h=A_-$, respectively. As $1/N$ goes to zero, $C_x^{(N)}$ and $m_z^{(N)}$
decrease for $h<A_-$ ($h=0.145, 0.155$) but increase for $h>A_-$ ($h=0.175,0.185$), indicating 
the discontinuity of $C_x^\infty$ and $m_z^\infty$.
Panels~(b) and~(d) plot $C_x^{(N)}$ and $m_z^{(N)}$ in the
vicinity of $h=A_+$, respectively. Again, we see the bifurcation as $1/N\to0$, which indicates
the discontinuity of $C_x^\infty$ and $m_z^\infty$.
Fig.~\ref{fig:p4scaling}(e) and fig.~\ref{fig:p4scaling}(f) plot
$\Delta$ as a function of $N$ at $h=A_-$ and $A_+$, respectively.
The data fit the curves $e^{-0.0505 N}$ at $h=A_-$ or $e^{-0.072N}$ at $h=A_+$.
We conclude that the gap vanishes exponentially at the transition points.

At $p=4$, the DPTs are first-order at $h=A_\pm(\kappa)$
which are both located inside the tri-stable semiclassical phase.
For $h\leq A_-$ or $h\geq A_+$, the LSS is similar to the semiclassical state $\mathcal{S}$.
For $A_-<h<A_+$, the LSS has vanishing magnetization but finite variance
in the $x$-direction, which is distinguished from either $\mathcal{S}$ or $\mathcal{P}_\pm$
but is an equal-weight mixture of $\mathcal{P}_+$ and $\mathcal{P}_-$.

\section{\label{sec:finiteNdynamics} The real-time dynamics}

\begin{figure}[tbp]
\vspace{0.5cm}
\includegraphics[width=0.9\linewidth]{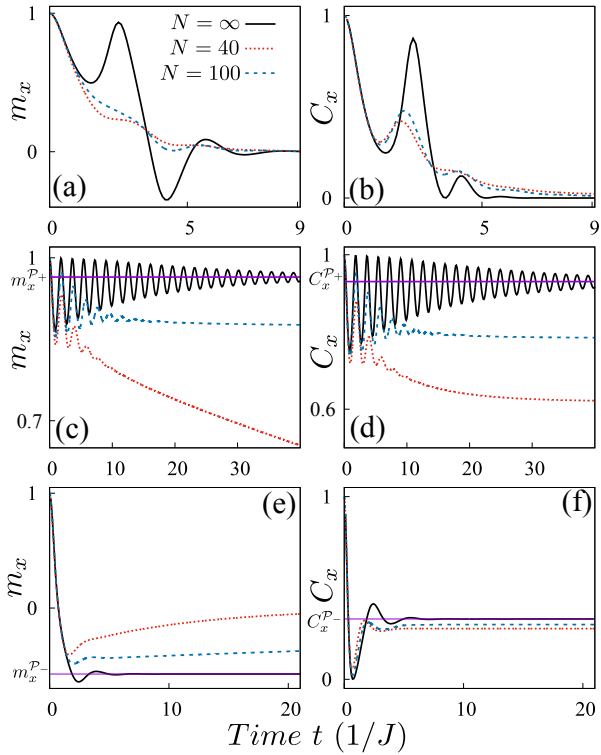}
\caption{(Color online) Transient dynamics of $m_x$ and $C_x$ at $p=2$.
Panels (a) and (b) are for $h/J=-0.5$, (c) and (d) for $h/J=0.1$, and (e) and (f) for $h/J=1.5$.
$\kappa=J$ is fixed. The spins are initialized in the positive $x$-direction
with $\left(\theta,\phi\right)=(\pi/2,0)$. Different line colors and types are for different $N$.
All the panels share the same legends.
The straight purple lines are $m_x=m_x^{\mathcal{P}_\pm}$ or $C_x=C_x^{\mathcal{P}_\pm}$.}\label{fig:dynamicsp2}
\end{figure}

In this section, we study the real-time dynamics of $m_\alpha$ and $C_x$ at finite $N$
and also in the semiclassical limit by solving Eq.~\eqref{eq:finiteNlind} or~\eqref{eq:meom}.

Fig.~\ref{fig:dynamicsp2} plots $m_x(t)$ and $C_x(t)$ in the transient regime for $p=2$
and different values of $(\kappa,h)$ whose positions in the parameter space are
marked by the black dots in fig.~\ref{fig:phase}(a). The initial condition is chosen to
$\theta=\pi/2$ and $\phi=0$, i.e., the spins are along the positive-$x$ direction.
The black solid lines are $m_x(t)$ and $C_x(t)$ in the semiclassical limit. As $t$ increases,
for $h/J=-0.5$, $0.1$ and $1.5$, $m_x(t)$ ($C_x(t)$) gradually approaches $m_x^{\mathcal{S}}$ ($C_x^{\mathcal{S}}$),
$m_x^{\mathcal{P}_+}$ ($C_x^{\mathcal{P}_+}$) and $m_x^{\mathcal{P}_-}$ ($C_x^{\mathcal{P}_-}$),
respectively. As $\kappa=J$ is fixed, $h/J=-0.5$, $0.1$ and $1.5$ are located in the $\{\mathcal{S}\}$-phase, $\{\mathcal{P}_+,
\mathcal{P}_-,\mathcal{S}\}$-phase and $\{\mathcal{P}_+,\mathcal{P}_-\}$-phase, respectively.
The relaxation of $m_x(t)$ and $C_x(t)$ in the semiclassical limit coincides with the analysis in Sec.~\ref{sec:semimethod}.
On the other hand, $m_x^{(N)}(t)$ or $C_x^{(N)}(t)$ at finite $N$ deviate significantly from their semiclassical limits.
Comparing the red ($N=40$) and blue ($N=100$) lines, we see that
the blue line is closer to the black solid line. Indeed, as $N$ goes to infinity,
$m_x^{(N)}(t)$ or $C_x^{(N)}(t)$ at arbitrary finite $t$ must
converge to their semiclassical limits, respectively. For small $t$ ($tJ< 1$),
$m_x^{(N)}(t)$ and $C_x^{(N)}(t)$ converge quickly with increasing $N$ and their values at $N=100$
are already indistinguishable from the semiclassical limits. But as $t$ increases,
a larger $N$ is needed for observing the convergence. For $tJ=10$, we see a clear difference between
$m_x^{(N)}(t)$ ($C_x^{(N)}(t)$) at $N=100$ and the semiclassical limits (see
fig.~\ref{fig:dynamicsp2}(c)-fig.~\ref{fig:dynamicsp2}(f)).

\begin{figure}[tbp]
\vspace{0.5cm}
\includegraphics[width=0.9\linewidth]{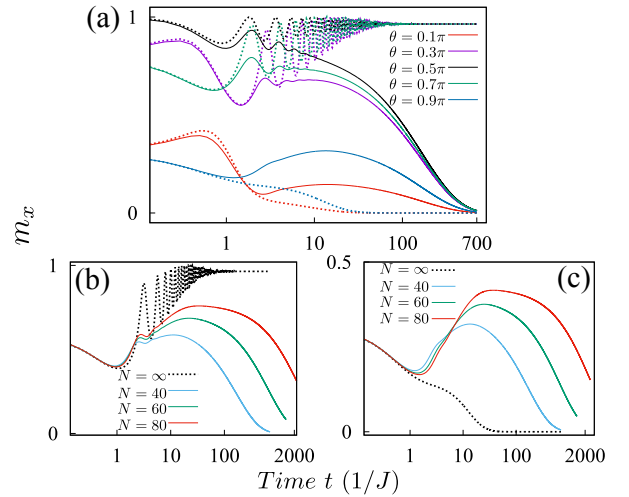}
\caption{(Color online) We plot $m_x(t)$ both at finite $N$ and in semiclassical limit with different initial conditions.
(a) Different line colors are for different $\theta$. For a given $\theta$,
the solid and dotted lines are for $N=40$ and $N\to\infty$, respectively.
(b) $m_x(t)$ at different $N$ with $\theta=0.8\pi$. (c) $m_x(t)$ at different $N$ with $\theta=0.9\pi$.}\label{fig:diss2}
\end{figure}

We choose $\left(\kappa,h\right)=\left(1,0.1\right)$ (in the $\left\{\mathcal{P}_+,
\mathcal{P}_-, \mathcal{S}\right\}$ phase) and plot $m_x(t)$ for
different initial conditions at much larger time scales in fig.~\ref{fig:diss2}(a). We fix $\phi=0$ and try
different values of $\theta$ (initial spins are aligned in the $xz$-plane with
positive $x$-component). Due to the $Z_2$ symmetry, $m_x^{(N)}(t)$ at finite $N$ always decays to zero
as $t\to\infty$ for whatever $\theta$. But in the semiclassical limit $N\to\infty$,
$\theta=0.1\pi$ or $0.9\pi$ lead to the steady magnetization $m_x^{\mathcal{S}}=0$,
while $\theta=0.3\pi$, $0.5\pi$ or $0.7\pi$ lead to a different steady
magnetization $m_x^{\mathcal{P}_+}\neq 0$. Since $m_x^{(N)}(t)$ at arbitrary $t$
goes to the semiclassical limit $m_x(t)$ as $N\to\infty$,
the relaxation time for $\theta=0.3\pi$, $0.5\pi$ or $0.7\pi$
must diverge with increasing $N$. We choose $\theta=0.8\pi$ (the semiclassical steady magnetization
is $m_x^{\mathcal{P}_+}$) to see how the relaxation time increases with $N$.
In fig.~\ref{fig:diss2}(b), the black dotted line is $m_x(t)$ in the limit $N\to\infty$.
The curve with $N=80$ is closer to the black dotted line than the curve with $N=40$.
As a consequence, the curve with $N=80$ starts to drop much later than the curve with
$N=40$. The relaxation time at $N=80$ is ten times larger than that at $N=40$.
In fig.~\ref{fig:diss2}(c), we choose $\theta=0.9\pi$ so that $m_x(t)$ at finite $N$ and
in the semiclassical limit both relax to zero. Under this initial condition,
$m^{(N)}_x(t)$ at finite $N$ goes up and down before it relaxes to zero,
and the relaxation time still diverges with increasing $N$.

In general, one can argue that the relaxation time must diverge in the multistable
semiclassical phase. Suppose there are two different semiclassical steady states, called $\mathcal{P}_+$
and $\mathcal{S}$ without loss of generality. At finite $N$, the system relaxes to
a unique steady state, which may be $\mathcal{P}_+$, $\mathcal{S}$
or different from both of them. Let us first suppose that the finite-size system relaxes to $\mathcal{P}_+$.
Now let us consider the initial condition $(\theta,\phi)$ under which the semiclassical steady state is
$\mathcal{S}$. For an arbitrarily large $T$, we can find a $N$ so that as the system's size is larger than $N$
it always keeps close to $\mathcal{S}$ for $t<T$.
But according to the above assumption, the system has to deviate from $\mathcal{S}$ and relax to
$\mathcal{P}_+$ in the long-time limit. This means that the relaxation time is larger than $T$.
But $T$ can be chosen arbitrarily large. Therefore, the relaxation time must be divergent as $N$ increases.
Similarly, we can argue that the relaxation time is divergent if the finite-size system relaxes
to $\mathcal{S}$ or some state different from both $\mathcal{P}_+$ and $\mathcal{S}$.
A divergent relaxation time corresponds to a vanishing Liouvillian gap.
We then guess that $\Delta$ should be zero within the whole multistable semiclassical phase.
This is consistent with what we observe in Sec.~\ref{sec:finiteN}. The gap certainly vanishes
at the DPTs, since the DPTs happen inside the multistable phase.

\begin{figure}[tbp]
\vspace{0.5cm}
\includegraphics[width=0.9\linewidth]{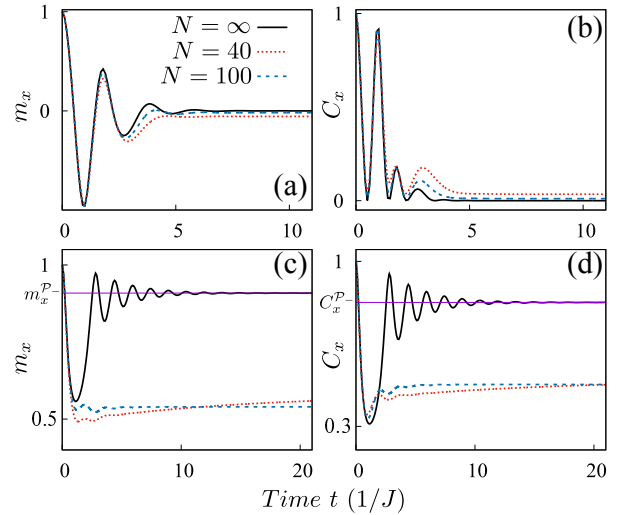}
\caption{(Color online) Transient dynamics of $m_x$ and $C_x$ at $p=3$.
Panels (a) and (b) are for $h/J=-1.6$, and (c) and (d) for $h/J=1.0$.
We set $\kappa=J$ and $\left(\theta,\phi\right)=(\pi/2,0)$.
All the panels share the same legends.
The straight purple lines are $m_x=m_x^{\mathcal{P}_-}$ or
$C_x=C_x^{\mathcal{P}_-}$.}\label{fig:dynamicsp3}
\end{figure}

The real-time dynamics at $p=3$ or $p=4$ exhibits similar features as $p=2$.
Fig.~\ref{fig:dynamicsp3} plots the transient dynamics of $m_x$ and $C_x$ for $p=3$
and different values of $(\kappa,h)$ that are located at the black dots in fig.~\ref{fig:phase}(b).
In fig.~\ref{fig:dynamicsp3}, The black solid lines are $m_x(t)$ and $C_x(t)$ in the semiclassical limit.
The parameter pair $\left(\kappa,h\right)=\left(1,-1.6\right)$ is located in the
$\{\mathcal{S}\}$-phase, in which both $m_x(t)$ and $C_x(t)$ relax towards zero
with increasing $t$ (see fig.~\ref{fig:dynamicsp3}(a) and~(b)).
While $\left(\kappa,h\right)=\left(1,1.0\right)$ is in the
$\{\mathcal{P}_-,\mathcal{S}\}$-phase, and we see that $m_x(t)$ ($C_x(t)$)
relaxes towards $m_x^{\mathcal{P}_-}$ ($C_x^{\mathcal{P}_-}$) (see fig.~\ref{fig:dynamicsp3}(c) and~(d)).
The blue and red dashed lines are $m_x^{(N)}(t)$ and $C_x^{(N)}(t)$
at finite $N$. Again, we see that $m_x^{(N)}(t)$ or $C_x^{(N)}(t)$ at small $t$
converge to their semiclassical limits as $N$ increases.
But the deviation of blue or red lines from the black solid line is significant
at large $t$, at which a larger $N$ is needed for observing the convergence.

\begin{figure}[tbp]
\vspace{0.5cm}
\includegraphics[width=0.9\linewidth]{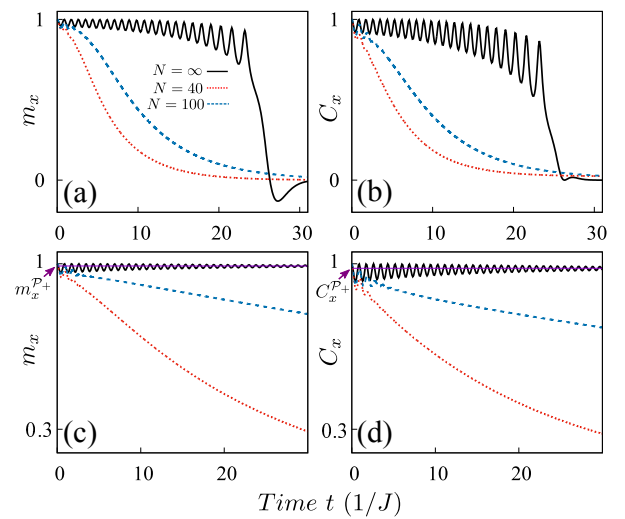}
\caption{
(Color online) Transient dynamics of $m_x$ and $C_x$ at $p=4$.
Panels (a) and (b) are for $h/J=-0.2$, and (c) and (d) for $h/J=0.2$.
We set $\kappa=J$ and $\left(\theta,\phi\right)=(\pi/2,0)$.
All the panels share the same legends.
The straight purple lines are $m_x=m_x^{\mathcal{P}_+}$ or $C_x=C_x^{\mathcal{P}_+}$.}\label{fig:dynamicsp4}
\end{figure}

Fig.~\ref{fig:dynamicsp4} plots the transient dynamics of $m_x$ and $C_x$
for $p=4$ and different values of $(\kappa,h)$ that are located at the black dots in fig.~\ref{fig:phase}(c).
In the semiclassical limit (black solid line), $m_x(t)$ ($C_x(t)$) relaxes to
either zero or $m_x^{\mathcal{P}_+}$ ($C_x^{\mathcal{P}_+}$), depending
on the value of $h$. The red and blue lines are for $N=40$ and $100$,
respectively. For $h/J=0.2$ (panels~(c) and~(d)), the blue or red lines significantly
deviate from the semiclassical prediction at large $t$.

\section{\label{sec:sum}Conclusions}

In this paper we studied a fully-connected $p$-spin model subject to a collective dissipation and in the presence of a Zeeman 
field. The model generalizes to generic $p$, the system studied
in references~[\onlinecite{Iemini17}] and~[\onlinecite{Hannukainen17}].
The effect of dissipation is to polarize  the spins along the $z$-direction
while the interaction between the spins favors the
alignment in the $x$-direction. The competition between the two
effects results in a complex phase diagram that depends strongly 
on the  value of $p$.

In the thermodynamic limit it is possible to derive the steady-state phase diagram through a semi-classical approach. 
Depending on the value of the couplings different steady states, including $\mathcal{S}$ and $\mathcal{P}_\pm$, 
appear as summarized in fig.~\ref{fig:phase} and table~\ref{tab}.
Within the semiclassical analysis, multi-stable regions  appear, with 
the steady-state magnetization in thermodynamic limit jumping between that of $\mathcal{S}$ and $\mathcal{P}_\pm$.
The  steady state with finite number of spins is unique, being independent of the initial condition. 
For even values of $p$, the system has a $Z_2$ symmetry, which guarantees the magnetization in the $x$-direction to be 
zero. For $p\geq 4$, $\mathcal{S}$ and $\mathcal{P}_\pm$ are distinguished from each other in the whole tri-stable region.
At the DPTs, the magnetization is discontinuous, jumping between zero (the value in $\mathcal{S}$)
and finite (the value in $\mathcal{P}_\pm$). The DPTs are first-order.
The case $p=2$ is however special. Except a first-order DPT happens inside the tri-stable region,
there also exists a bistable region in which $\mathcal{P}_\pm$
are stable but $\mathcal{S}$ is not. With increasing Zeeman field, $\mathcal{P}_\pm$
move towards $\mathcal{S}$. They meet each other at a critical point,
where the magnetization variance vanishes continuously, indicating a continuous DPT.
On the opposite, when $p$ is odd, the $Z_2$ symmetry is explicitly broken. First-order DPTs
happen inside the bistable region in which both $\mathcal{S}$ and $\mathcal{P}_-$
are stable. The magnetizations in both the $x$- and $z$-directions are
discontinuous at the DPTs, where they jump between the values in $\mathcal{S}$
and $\mathcal{P}_-$.

In order to further understand the properties of the transition, we analyzed the finite-scaling properties in the 
transition regions. To this end, we studied the behavior of the Liouvillian gap as a function of the number of spins.
Interesting scaling behavior was found at the continuous DPT, where both the magnetization variance and Liouvillian gap vanish 
with increasing number of spins according to a power law. On the opposite, the Liouvillian gap was found to vanish
exponentially at the first-order DPT.

\section*{Acknowledgement}
Pei Wang is supported by NSFC under Grant Nos.~11774315 and~11835011,
and by the Junior Associates program of the Abdus Salam International Center for Theoretical Physics.
Rosario Fazio acknowledges partial financial support from the Google Quantum Research Award.
R. F. research has been conducted within the framework of the Trieste Institute for
Theoretical Quantum Technologies (TQT).

\end{document}